\documentclass[twocolumn,showpacs,preprintnumbers]{revtex4}
\usepackage{graphicx}
\usepackage{dcolumn}
\usepackage{bm}
\usepackage{amsmath}

\begin{document}

\title{Recombination of H$_3^+$ Ions in the Afterglow of a He-Ar-H$_2$ Plasma}
\author{J. Glosik$^\dagger$, I. Korolov$^\dagger$, R. Plasil$^\dagger$, O. Novotny$^\dagger$, T. Kotrik$^\dagger$, P. Hlavenka$^\dagger$, J. Varju$^\dagger$,\\
Chris H.~Greene$^\ddagger$, V.~Kokoouline$^*$, and I.~A.~Mikhailov$^{*,**}$}
\affiliation{$^\dagger$Mathematics and Physics Faculty, Charles University in Prague, Prague 8, Czech Republic;\\
$^\ddagger$Department of Physics and JILA, University of Colorado, Boulder, Colorado 80309-0440, USA; \\
$^*$Department of Physics, University of Central Florida, Orlando, Florida 32816, USA;\\
$^{**}$Petersburg Nuclear Physics Institute, Gatchina, St. Petersburg 188300, Russia.}

\date{\today}

\begin{abstract}
Recombination of H$_3^+$ with electrons was studied in a low temperature plasma in helium. The plasma recombination rate is driven by two body, H$_3^+ +$ e$^-$, and three-body, H$_3^+ +$ e$^-$+ He, processes with the rate coefficients $7.5\times10^{-8}\,$cm$^{3}$s$^{-1}$ and $2.8\times10^{-25}\,$cm$^{6}$s$^{-1}$ correspondingly at 260~K. The two-body rate coefficient is in excellent agreement with results from storage ring experiments and theoretical calculations. We suggest that the three-body recombination involves formation of highly excited Rydberg neutral H$_3$ followed by an $l$- or $m$- changing collision with He. Plasma electron spectroscopy indicates the presence of H$_3$. 
\end{abstract}


\maketitle

Recombination of H$_3^+$ with electrons has been the subject of many experimental and theoretical studies for nearly 50 years. 
Because of the high hydrogen abundance in the universe, H$_3^+$ plays a key role in interstellar molecular clouds and planetary atmospheres \cite{oka00}. It is also an important ion in many laboratory hydrogen-containing plasmas and in technological applications, including cold regions of fusion devices. Despite enormous effort, the process of H$_3^+$ recombination is still controversial (see the critical discussion in Ref. \cite{johnsen05}). 

Early theoretical studies suggested that the H$_3^+$ recombination rate is slow, although experiment gave a much higher recombination rate. Inclusion of non-Born-Oppenheimer Jahn-Teller coupling significantly improved the theory and  \cite{kokoouline01,kokoouline03} gave a high rate of the dissociative recombination (DR) with the rate coefficient (RC) $\alpha=7.8\times10^{-8}\,$cm$^{3}$s$^{-1}$ at 300~K. Recent storage ring experiments \cite{mccall03,kreckel05} with cold ion sources obtained a DR rate for 300~K in good agreement with theory. Discrepancies remain just in connection with para and ortho H$_{3}^{+}$ DR. The disagreement with plasma experiments has not been solved until now \cite{Plasil02,Glosik}.

Essentially two different types of the experiments have measured the DR RCs: Colliding beam experiments \cite{Wolf}, where binary cross sections are measured, and afterglow experiments \cite{Plasil02} where one monitors the decay of the H$_3^+$-dominated plasma and the recombination RC $\alpha(T)$ versus temperature is obtained. The essential difference is that in the beam experiments only binary collisions are important, whereas in a plasma many-body processes are possible and can play an important role. 

In this article we present new experimental data for recombination in an H$_3^+$-dominated plasma. The new results and the analysis reveal the importance of three-body processes in plasma recombination and finally reconcile the theoretical and experimental storage ring results with the plasma experiments.

\begin{figure}[ht]\includegraphics[width=7cm]{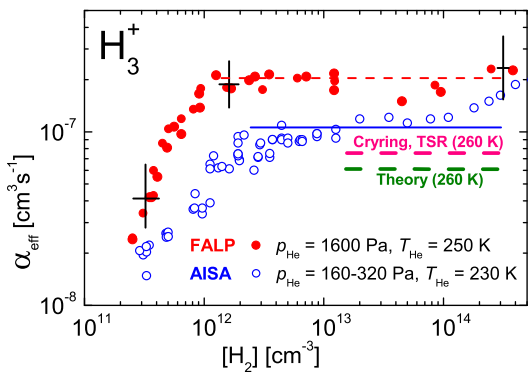}\caption{\label{fig1} The variation of recombination RC with H$_2$ density for two different He pressures. The data were obtained in AISA \cite{Plasil02,Glosik} and FALP experiments. The theoretical value for the binary RC is indicated \cite{kokoouline01,kokoouline03} , as is the current value from storage-ring experiments \cite{mccall03,kreckel05,Wolf}.}
\end{figure}

The first experiment discussed here is the advance integrated stationary afterglow experiment (AISA), where a plasma was generated by pulsed microwave discharge in a He/Ar/H$_{2}$ mixture \cite{Plasil02,Glosik}. The presence of He/Ar/H$_2$ is essential because H$_3^+$ is formed in the sequence of ion-molecule reactions \cite{Plasil02,Glosik}. After the microwaves shut off, the decay of the cold H$_3^+$ dominated plasma is monitored over 60~ms with a Langmuir probe. From this plasma decays, recombination RCs as small as $5\times10^{-9}\,$cm$^{3}$s$^{-1}$ can be measured with AISA. In AISA we studied the dependence of recombination RC on the partial pressure of H$_2$. Examples of the data obtained are plotted in Fig. \ref{fig1}. The plasma recombination RC $\alpha_{\rm eff}$ is extracted under the assumptions that the plasma is quasi-neutral and that the decay is dominated by processes described by the balance equation:
\begin{equation}\frac{ d n_e}{ dt}=-\alpha_{\rm eff}n_e n_+=-\alpha_{\rm eff}n_e^2\,,\end{equation}where $n_{e}$ and $n_{+}$ are the electron and ion densities, respectively. The AISA data plotted in Fig. \ref{fig1} show that at low hydrogen densities, $\alpha_{\rm eff}$ increases with [H$_2$] until it reaches ``saturation'' at [H$_2$] $\sim2\times10^{12}\,$cm$^{-3}$.

To verify that in our plasma experiments ions other than H$_3^+$ do not play an important role in recombination (see discussion in \cite{Smith}) we built another stationary afterglow experiment -- Test Discharge Tube equipped with a near-infrared Cavity Ring Down Spectrometer, TDT-CRDS \cite{Macko04a}. In this experiment the decay of H$_3^+(v=0)$ was monitored during the afterglow (after microwaves shut off) and $\alpha_{\rm eff}$ was obtained. From the Doppler broadening of the absorption line the kinetic temperature of ions was determined \cite{Macko04b}. High electron (and ion) density was used in the TDT-CRDS experiment in order to have sufficient signal-to-noise ratio. The results of TDT-CRDS obtained in the ``saturated'' region agree well with the AISA  data.

\begin{figure}[ht]\includegraphics[width=7cm]{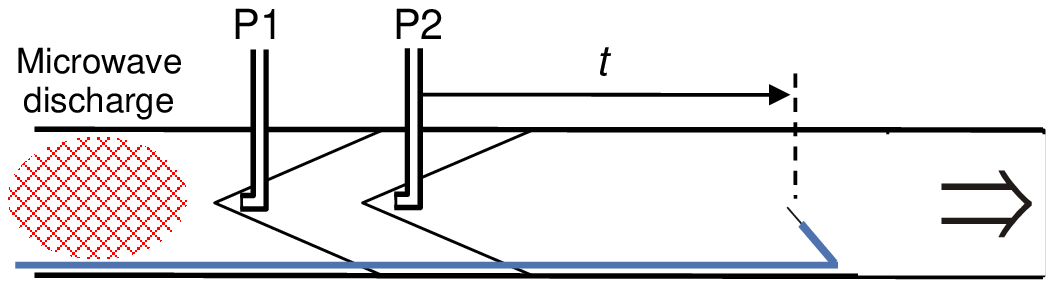}\caption{\label{fig2} Principle of the FALP method. Buffer gas (He) flows from the discharge region towards a pump (right side). Reactants are added via ports P$_1$ and P$_{2}$. The relation between time $(t)$ and position is given by buffer gas velocity.}
\end{figure}

To obtain even higher control over the plasma parameters in a recombination-dominated afterglow we have adapted a flowing afterglow apparatus (FALP variant \cite{Glosik03}, see Fig. \ref{fig2}) to study the slow recombination processes with RCs down to $10^{-8}\,$cm$^{3}$s$^{-1}$. In this FALP, the plasma is formed upstream in the microwave discharge. To remove metastable helium by Penning ionization, Ar is added through port P$_{1}$. In the He/Ar mixture an Ar$^+$ dominated plasma is formed \cite{Plasil02,Glosik03} and carried along the flow tube for 30~ms. When the plasma is already cold, H$_2$ is introduced via port P$_2$, which forms an H$_3^+$ dominated plasma \cite{Plasil02}. The decay of the H$_3^+$ dominated plasma is monitored by Langmuir probe along the flow tube for another 60~ms. The coefficients $\alpha_{\rm eff}$ obtained by FALP for different [H$_2$] at 1600~Pa and 250~K are plotted in Fig. \ref{fig1}. Again, a dependence of $\alpha_{\rm eff}$ on [H$_2$] was observed. It is also clear that $\alpha_{\rm eff}$ depends on He pressure, suggesting that the recombination in a plasma is not just a binary process. Note the small increase of $\alpha_{\rm eff}$ at [H$_2$] approaching $2\times10^{14}\,$cm$^{-3}$, this is due to the formation of H$_5^+$ ions in three-body association reaction, H$_3^+$ + H$_2$ + He $\to$ H$_5^+$ + He, and their immediate recombination with electrons. This process was studied separately using FALP and TDT-CRDS and we obtained very clear results and agreement with thermodynamic data on H$_{5}^{+}$ formation \cite{Glosik03}. On the base of these studies we can conclude, that influence of H$_5^+$ formation on measured $\alpha_{\rm eff}$ can be excluded at 260~K for [H$_2$] $\leq 10^{14}\,$cm$^{-3}$.

\begin{figure}[ht]\includegraphics[width=7cm]{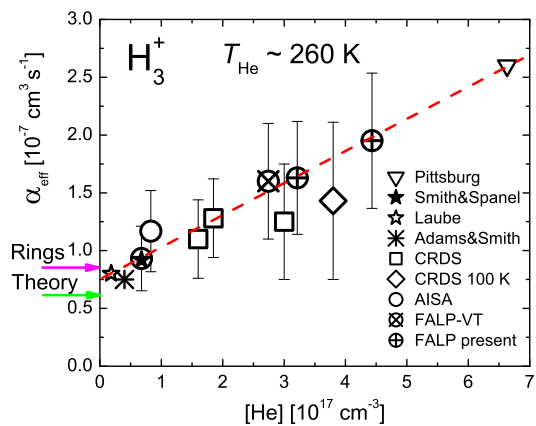}\caption{\label{fig3} The dependence of effective recombination RC $\alpha_{\rm eff}$ on He number density. A straight line is plotted through the AISA and Pittsburgh data points.}
\end{figure}

{\bf Region of high H$_2$ density}. In order to understand the mechanism of recombination we concentrate first on the ``saturated'' region, i.e. on data obtained with $2\times10^{12}\,$cm$^{-3}\leq$ [H$_2$] $\leq 1\times10^{14}\,$cm$^{-3}$. In these conditions, $\alpha_{\rm eff}$ depends only on the He pressure. 
Fig.~\ref{fig3} plots the values of $\alpha_{\rm eff}$ obtained in AISA, FALP and TDT-CRDS experiments in the ``saturated'' region at 260~K for various He densities. Fig. \ref{fig3} also includes the theoretical RC \cite{santos07} and results obtained in other laboratories where a He/H$_2$ mixture was used; in cases where their data were obtained at different temperature (but close to 260~K) the corresponding value $\alpha_{\rm eff}$(260~K) was calculated assuming a $T_{e}^{-0.5}$ dependence. The following observations are relevant to interpreting the plotted data:\\
({\bf i}) Pittsburgh group, stationary afterglow \cite{Leu73}. The measurement was in the saturation region at 18.7 Torr and 300~K. $\alpha_{\rm eff}= 2.3\times10^{-7}\,$cm$^{3}$s$^{-1}$ (Fig. 2 of Ref. \cite{Leu73}). 
(\textbf{ii}) Smith and Spanel (Fig.~4 of Ref. \cite{Smith}). At the beginning of the decay $\alpha_{\rm eff}$(300~K) $\sim 8\times10^{-8}\,$cm$^{3}$s$^{-1}$ at 2 Torr. For sufficiently long decay time the recombination coefficient becomes very small.
({\bf iii}) Laube {\it et al.}, FALP at 0.5 Torr. $\alpha_{\rm eff}$(300~K) $=7.8\times10^{-8}\,$cm$^{3}$s$^{-1}$ (abstract and Fig.~3 of Ref. \cite{Laube98}). 
(\textbf{iv}) Adams and Smith, FALP. Fig~2. of Ref. \cite{Adams84} shows the decay curve with a fast decay at the beginning and a slower decay towards the end of the flow tube. Interpretation for this behavior will be discussed elsewhere. Here, we use the rate constant of the fast decay, $\alpha_{\rm eff}$(300~K) $=7\times10^{-8}\,$cm$^{3}$s$^{-1}$ at 1.2 Torr. 
(\textbf{v}) Storage ring with cold ion sources \cite{mccall03,kreckel05}.
(\textbf{vi}) Our lab, TDT-CRDS, present data, and data from Refs. \cite{Macko04a,Macko04b,Plasil05}. These are the only measurements made in He/Ar/H$_2$ mixture using absorption spectroscopy. The decay was measured in relatively early afterglow, at $50-300$~$\mu$s and electron densities $n_{e0}\sim 5\times10^{11}$cm$^{-3}$.  
(\textbf{vii}) Our lab - stationary afterglow, AISA and VT-AISA, present data and data from Refs. \cite{Plasil02,Glosik}.
(\textbf{viii}) Our lab -- FALP (two different constructions), present study and data from Refs. \cite{Plasil02,Pysanenko}. 
(\textbf{ix}) Theoretical value \cite{santos07}.
Experimental data from 10 very different plasma experiments measured over 24 years suggest that effective $\alpha_{\rm eff}$ depends linearly on the He density
\begin{equation}\label{eq:linear_He}\alpha_{\rm eff}=\alpha_{\rm eff}^0+K_{\rm He}[\rm He]
\end{equation}with the coefficient of three-body recombination $K_{\rm He}\approx 2.8\times 10^{-25}\,$cm$^6$s$^{-1}$. An extrapolation for [He]$\to$0 gives the value $\alpha^0_{\rm eff}(260\,{\rm K})=7.5\times 10^{-8}\,$cm$^{3}$s$^{-1}$. This value agrees well with the H$_3^+$ DR RC obtained in several storage ring experiments \cite{mccall03,kreckel05,Wolf} and theory \cite{santos07}.

The observed linear dependence of $\alpha_{\rm eff}$ as a function of the helium density can be attributed to an additional channel of recombination of H$_3^+$, which is effective only in presence of He and can be called He-assisted recombination:
\begin{equation}\label{eq:He_assisted_DR}
\rm{H}_{3}^{+} + {\rm e}^{-} + \rm{He}\ \substack{K_{\rm He}\\ \longrightarrow} \ \left\{ \begin{array}{ll}
{\rm H_{2}+H+He \ \ \ \ or}\\{\rm H+H+H+He\ \ \ \ or}\\{\mathrm{\tilde{H}}_{3}^{\ast }\rm +He}\end{array}\right.\end{equation}with the three-body RC \textit{K}$_{\rm{He}}$. The first two channels in Eq. \ref{eq:He_assisted_DR} can be called He-assisted DR. The molecules $\mathrm{\tilde{H}}_{3}^{\ast }$ have a finite lifetime and can be relatively stable only in an excited electronic state. Such metastable molecules play am important role at low densities of background H$_2$ gas (see below). Previously, three-body processes have been observed and described by Thomson and later Bates \cite{Bates}. Typical values of three-body RCs for He as a third body are $10^{-27}\,$cm$^{6}$s$^{-1}$ (see refs. \cite{Cao,Bates} and references therein). Processes with such RCs can contribute only at pressures higher than 1 atm. The process we observed is more efficient by factor of 100. It is however possible that the three-body process of Eq. (\ref{eq:He_assisted_DR}) proceeds through an intermediate state of H$_{3}^{*}$ formed during binary H$_{3}^{+} + {\rm e}^{-}$ collisions. If there is no external perturber (such as He or H$_{2}$), the lifetime of such H$_{3}^{*}$ is mainly determined by autoionization: the autoionization probability is much larger than the predissociation. Typical lifetimes of electronic \textit{p}-states of H$_{3}^{*}$ (i.e., those most likely to be formed in H$_{3}^{+}$ + e$^{-}$ collisions) are $\tau_{a}=10-100$~ps.
If, during this time, H$_{3}^{*}$ collides with a He atom followed by a change of the electronic state of H$_{3}^{*}$, (say, $l$-changing H$_3^*$ + He collision) the new state should have a much longer lifetime with respect to autoionization. These states may be stable or not with respect to predissociation. We use the notation $\mathrm{\tilde{H}}_{3}^{\ast}$
(in contrast to H$_3^*$) in Eq. \ref{eq:He_assisted_DR} and below to label the states stable with respect to predissociation and autoionization. It is possible to estimate the three-body RC $\tilde k^{\ast}$ for formation of such states.

The number $\tilde N^\ast$ of $\mathrm{\tilde{H}}_{3}^{\ast }$ molecules created in 1~cm$^{3}$ per 1 second is $k_l$[He][H$_{3}^{*}$], where $k_l$ is the RC for the \textit{l}-changing collisions. Assuming that H$_{3}^{*}$ are constantly being created in H$_{3}^{+}+ {\rm e}^-$ collisions and destroyed by autoionization, their density [H$_{3}^{*}$] can be estimated as [H$_{3}^{*}$] = [H$_{3}^{+}$]$P$, where \textit{P} is the fraction of time that an ion H$_3^+$ is being in the ``form'' of H$_{3}^{*}$. Thus, $P=\tau_a/\tau_c$, where $\tau_c=1/(\alpha^*n_e)$ is the time between H$_{3}^{+}+{\rm e}^{-}$ collisions, $\alpha^*$ is the RC for H$_3^*$ formation. Therefore, $\tilde N^{\ast}=k_l$[He][H$_{3}^{+}$]$\tau_a \alpha^*n_e$, and the three-body RC is estimated as $\tilde k^{\ast}=k_l \tau_a \alpha^*$.

The RC $\alpha^*$ is larger than the DR RC $\alpha_{\rm DR} =7.5\times10^{-8}\,$cm$^{3}$s$^{-1}$ ($T=260$~K) \cite{kokoouline03} by about a factor 10, i.e. $\alpha^*\sim 8\times10^{-7}\,$cm$^{3}$s$^{-1}$. The RC $k_l$ for the \textit{l}-changing collisions can be estimated from experimental and theoretical data for Na$^*$+He \textit{l}-changing collisions \cite{Hickman}: $k_l\sim 2.3\times10^{-8}\,$cm$^{3}$s$^{-1}$. If we take $\tau_a=20$~ps, the estimated value for $\tilde k^{\ast}$ is $4\times10^{-25}\,$cm$^{6}$s$^{-1}$.

{\bf Region of low H$_2$ density.} When the density of the H$_2$ background gas is smaller than $2\times 10^{12}\,$cm$^{-3}$ the AISA and FALP experiments show almost the linear dependence of effective $\alpha_{\rm eff}$ on [H$_2$]. This can be explained by the presence of long-living states of the neutral H$_3$ molecule that are formed in the decaying plasma (the third channel in Eq. \ref{eq:He_assisted_DR}) and can only be destroyed in collisions with H$_2$ molecules. Such metastable molecules $\mathrm{\tilde{H}}_{3}^{\ast }$ can be formed in the $l$-changing collisions H$_3^*$ + He discussed above. Because they can only be destroyed by H$_2$, at low H$_2$ densities such $\mathrm{\tilde{H}}_{3}^{\ast }$
molecules play a role of the ``storage reservoir'' for the H$_3^+$ ions. Slow release of  the H$_3^+$ ions from the $\mathrm{\tilde{H}}_{3}^{\ast }$
molecules makes measured effective $\alpha_{\rm eff}$ to be dependent on the H$_2$ density.

Our calculation shows that the molecules H$_3^*$ (having relatively short lifetime $10-100$~ps) are formed mostly in highly excited Rydberg $np$ states with $n\sim 20-70$. This is because the second ionization  threshold for para-H$_3^+$ is 170~cm$^{-1}$ above the lowest ionization threshold. This difference happens to coincide approximately with the temperature of the experiment $T=260$~K. At energies below the second ionization threshold, the H$_3$ system has an infinite number of $p$-state resonances with widths comparable to the splitting between resonances. The probability of capture to one of these resonances is comparable to unity \cite{kokoouline03}. Thus, if  $\mathrm{\tilde{H}}_{3}^{\ast }$
is formed from H$_3^*$ in $l$-changing collisions with He, they are formed in highly excited Rydberg states that should have very long lifetimes since the lifetime scales as $n^3$.
In fact, at even lower electron temperatures $\lesssim100$~K, the binary DR RC is predicted to be significantly smaller \cite{santos07} for ortho-H$_3^+$ than for para-H$_3^+$. This should partially compensate for the ``storage effect'' at lower temperatures.
The present experimental data suggest that such metastable molecules are indeed present in the decaying plasma, as is shown below.

\begin{figure}[ht]\includegraphics[width=7cm]{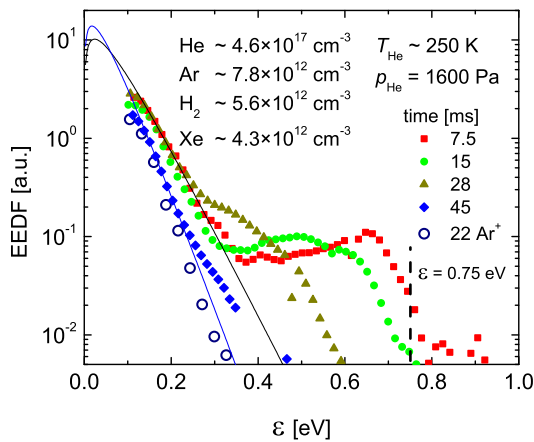}\caption{\label{fig4} EEDF measured in FALP experiment in He/Ar/H$_2$/Xe decaying plasma: EEDFs in several times of the plasma decay are depicted. The data obtained have been fitted to a Maxwellian distribution (full lines) and the EEDF are normalized. For comparison we also show the EEDF (open circles) measured in an Ar$^+$ dominated plasma without addition of the H$_2$/Xe mixture.}
\end{figure}

Thus, the presence of He and H$_2$ influences the overall process of plasma recombination: When H$_3^+$ collides with an electron, a helium atom can speed up the dissociation of H$_3^*$ or help to create metastable molecules 
$\mathrm{\tilde{H}}_{3}^{\ast }$.
If there are many H$_2$ diatomic molecules such metastable molecules are destroyed. It is the H$_2$-``saturated'' region. In this region, the recombination RC is described by Eq.~\eqref{eq:linear_He}. When the density of H$_2$ is small, the 
$\mathrm{\tilde{H}}_{3}^{\ast }$
molecules are accumulated and play a role of a temporary storage for H$_3^+$ ions. For this region the effective recombination rate $\alpha_{\rm eff}$ depends on  [H$_2$].

If $\mathrm{\tilde{H}}_{3}^{\ast }$
exist and if collisions with H$_2$ influence the process of recombination, then if another reactant will be added with H$_2$ we should observe the reaction. There are not many possibilities; if $\mathrm{\tilde{H}}_{3}^{\ast }$
has to be also produced Xe is probably the only choice. After addition of Xe a new channel is open:
$\mathrm{\tilde{H}}_{3}^{\ast }+ \rm {Xe} \rightarrow \rm {XeH}^+ + {\rm e}_{\rm {fast}}^- + \rm {H}_2$. If $\mathrm{\tilde{H}}_{3}^{\ast }$
is highly excited, with energy close to ionization, then this reaction is exothermic and the ejected electron can have energy up to 0.75~eV. Fast electrons with energy 0.75~eV produced in cold plasma with electron temperature $T_{e}\sim 0.03$~eV can be in principle detected by measuring the electron energy distribution function (EEDF) \cite{Glosik99,Arslanbekov}. This method is named plasma electron spectroscopy (PES) \cite{Kolokolov,Korolov}. In the experiment the discharge was in He, then Ar was added via port P$_1$, and 30~ms later, Xe and H$_2$ were added through port P$_2$ considered as $t=0$. The densities of H$_2$ and Xe were made similar, in order to have comparable probabilities of reaction with $\mathrm{\tilde{H}}_{3}^{\ast }$.

Measured EEDFs in He/Ar/H$_2$/Xe afterglow plasma are depicted in Fig. \ref{fig4}. The measured EEDFs clearly indicate a production of fast electrons with energy $\varepsilon\sim 0.75$~eV in the plasma, as expected for the Xe reaction process. The method is not yet sufficiently elaborated to account for potential superelastic collisions of electrons with internally excited ions (e.g. XeH$^+$) and to give quantitative results. Qualitatively, the measured EEDF evolution indicates that at low decay time, when the recombination is fast, the production of energetic electrons is also fast; at a longer decay time the production becomes slower and EEDF approaches a Maxwellian shape.

Analysis of the present and several previous experiments involving an H$_3^+$-dominated plasma provides firm evidence that the rate of plasma recombination strongly depends on the density of the background helium.
More importantly, the analysis suggests that at the limit of zero He density, the thermal RC for binary recombination in all considered plasma experiments is in agreement with the recent storage ring experiments and with theory. Thus, the 50-year old problem of disagreement for the binary DR RC at low energies between different experiments and theory appears to be resolved.
Finally, the present experimental data suggests that long-lived $\mathrm{\tilde{H}}_{3}^{\ast }$ molecules exist in the H$_3^+$-dominated plasma and play an important role in the plasma recombination.
Acknowledgements: This work is a part of the research plan MSM 0021620834 financed by the Ministry of Education of the Czech Republic and partly was supported by GACR (202/05/P095, 205/05/0390, 202/03/H162, 202/07/0495) by GAUK 53607 and GAUK 124707 and by the National Science Foundation, Grants No. PHY-0427460, PHY-0427376.

\end{document}